\documentclass[reqno]{article}

\usepackage{graphicx}
\usepackage{amsmath,amsthm,amssymb}

\chardef\bslash=`\\ % p. 424, TeXbook
\theoremstyle{definition}

\theoremstyle{remark}

\newcommand{\eval}[2][\right]{\relax
  \ifx#1\right\relax \left.\fi#2#1\rvert}

\begin{document}
\title{\bf{Self-similar spherically symmetric wave maps coupled to gravity}}

\author{Piotr Bizo\'n\footnotemark[1]{}\,\; and
  Arthur Wasserman\footnotemark[2]{}\\
  \footnotemark[1]{} \small{\textit{Institute of Physics,
   Jagellonian University, Krak\'ow, Poland}}\\
   \footnotemark[2]{} \small{\textit{Department of Mathematics,
   University of Michigan, Ann Arbor, Michigan}}}
\maketitle
\begin{abstract}
\noindent We investigate spherically symmetric continuously
self-similar (CSS) solutions in the $SU(2)$ sigma model coupled to
gravity. Using mixed numerical and analytical methods, we provide
evidence for the existence (for small coupling) of a countable
family of regular CSS solutions. This fact is argued to have
important implications for the ongoing studies of black hole
formation in the model.
\end{abstract}

\section{Introduction}
Self-similar solutions of  Einstein's equations have been
extensively studied in general relativity (see~\cite{carr} for the
recent review and vast bibliography of this subject). They are
interesting for several reasons. First, under the assumption of
self-similarity Einstein's equations simplify considerably and
sometimes (in spherical symmetry, for instance) reduce
 to a system of ordinary differential equations. This enables one
 to study them using dynamical systems methods and in some cases it
 is even possible to find explicit solutions. Second, self-similar
 solutions exhibit a well-defined singular behaviour at the
 blow-up. This makes them relevant in the studies of singularity
 formation; in fact, most of the examples of naked singularities
 in the literature
 involve self-similar solutions. Third, in some situations solutions of
 the Cauchy problem
 starting from generic initial data evolve asymptotically to
 self-similar form, in other words, self-similar solutions can
 play the role of attractors.
 Finally, the recent surge of interest in self-similar solutions
 is due to the fact that they appear
 as critical solutions at the threshold for black hole formation
 in gravitational collapse.

 Such reasons motivated also this paper
which, in a sense, is an extension of~\cite{ja} where self-similar
wave maps from
 Minkowski spacetime into the 3-sphere were studied. It was proved
 in~\cite{ja} that there exists a countable family of
 regular (by regularity we mean analyticity below the Cauchy horizon of the
  singularity)
 CSS solutions, labeled by a nonnegative integer $n$ (a nodal number).
  It was also
 shown that the $n$th solution has exactly $n$ unstable modes. The
 role of these solutions in the dynamical evolution was
 investigated in~\cite{my} where it was shown that: (i) the $n=0$ solution
 determines a universal asymptotics of singularity formation, (ii)
 the $n=1$ solution appears as a  critical solution at the threshold for singularity
 formation (see also~\cite{hli}). Here, we generalize~\cite{ja} by turning on
  gravity, that is,
  we consider wave maps with the 3-sphere as the target and a
  domain manifold  which is not Minkowski but a spacetime
 satisfying the Einstein equations with the $SU(2)$ sigma field as
 the source.
 The coupling of gravity is parametrized by a dimensionless
 coupling constant $\alpha$ (so that $\alpha=0$ corresponds to
 gravity being turned off). Since the coupling does not break scale
 invariance, it is natural to ask if the $\alpha=0$ CSS
 solutions constructed in~\cite{ja}, persist for nonzero $\alpha$.
A positive answer to this question is the main result of our
paper. More precisely, we show that the $n$th CSS solution which
is analytic within the past light cone of the singularity (in what
follows we will refer to the past and future light cones of the
singularity as to the past and future self-similarity horizons
(SSH)), persists up to $\alpha=1/2$. However, this solution is
regular (that is, analytic up to the future SSH) only for
$\alpha<\alpha_n$, where $\{\alpha_n\}$ is an increasing sequence
bounded above by $1/2$. For $\alpha_n<\alpha<1/2$, the past SSH of
the $n$th solution is surrounded by a spacelike apparent horizon
beyond which the solution cannot be smoothly continued. Although
we have not studied linear stability, it seems plausible that the
stability properties of solutions for small $\alpha$ are the same
as for $\alpha=0$. Assuming this, the $n=0$ solution (for
$\alpha<\alpha_0 \approx 0.0688$) is a generic naked singularity,
while the $n=1$ solution (for $\alpha<\alpha_1 \approx 0.1518$) is
a codimension-one naked singularity and the candidate for a
critical solution.

The CSS solutions constructed here should be relevant for the
character of type II critical collapse in the model which is
currently being investigated by Husa, Lechner, P\"urrer,
Thornburg, and Aichelburg (HLPTA). Their first report of
progress~\cite{husa1} shows discretely self-similar (DSS)
behaviour at the threshold for black hole formation for $\alpha
\geq 0.18$ and indicates that DSS disappears for small $\alpha$.
It is natural to expect that for $\alpha<\alpha_1$ the $n=1$ CSS
solution takes over as a critical solution. Preliminary results of
HLPTA seem to confirm this expectation~\cite{husa2}. A mechanism
of the DSS/CSS changeover is not yet understood, however. We
anticipate that the standard type II CSS critical behaviour will
be observed only in the range $\alpha_0 <\alpha<\alpha_1$. For
$\alpha<\alpha_0$ we expect qualitatively the same behaviour as in
the flat spacetime, where the $n=1$ solution sits at the threshold
for the $n=0$ naked singularity formation.

The rest of the paper is organized as follows. In section~2 we
define the model and derive the field equations. In section~3 we
construct a countable family of CSS solutions which are analytic
between the center and the past SSH. An extension of these
solutions beyond the past SSH is determined in section~4. In
section~5 we speculate about the role of regular CSS solutions in
the dynamical evolution.
\section{Field Equations}
Let $X: M \rightarrow N$ be a map from a spacetime $(M,g_{ab})$
into a Riemannian manifold $(N,G_{AB})$. Wave maps coupled to
gravity are defined as extrema of the action
\begin{equation}
S = \int_M \left(\frac{R}{16 \pi G}  + L_{WM}\right) dv_g
\end{equation}
with the Lagrangian density
\begin{equation}
  L_{WM} = -\frac{f^2_{\pi}}{2} g^{ab} \partial_a X^A \partial_b X^B
  G_{AB}.
\end{equation}
Here $G$ is Newton's constant and $f^2_{\pi}$ is the wave map
coupling constant. Remarkably, in the "physical" case of $3+1$
dimensional spacetime (to which our paper is confined), the
coupling of gravity does not break scale invariance but introduces
the dimensionless coupling constant $\alpha=4\pi G f^2_{\pi}$. The
field equations derived from (1) are the wave map equation
\begin{equation}
\square_g X^A + \Gamma_{BC}^A(X) \partial_aX^B \partial_bX^C
g^{ab}=0,
\end{equation}
where $\Gamma_{BC}^A(X)$ are the Christoffel symbols of the target
metric $G_{AB}$ and $\square_g$ is the d'Alembertian associated
with the metric $g_{ab}$, and  the Einstein equations
$R_{ab}-\frac{1}{2} g_{ab} R = 8 \pi G T_{ab}$ with the
stress-energy tensor
\begin{equation}
  T_{ab} = f^2_{\pi} \left(\partial_a X^A \partial_b X^B
  -\frac{1}{2} g_{ab}( g^{cd} \partial_c X^A \partial_d X^B)\right)
  G_{AB}.
\end{equation}
 The target
manifold is taken as the three-sphere $S^3$ with the standard
metric in polar coordinates $X^A=(F,\Theta,\Phi)$
\begin{equation}
G_{AB} dX^A dX^B = dF^2 + \sin^2{\!F} \:(d\Theta^2 +
\sin^2{\!\Theta}\: d\Phi^2).
\end{equation}
For the domain manifold we assume spherical symmetry and use
Schwarzschild coordinates
\begin{equation}
g_{ab}dx^a dx^b= -e^{-2 \delta} A\: dt^2 + A^{-1} dr^2 + r^2
(d\theta^2+\sin^2{\!\theta}\: d\phi^2),
\end{equation}
 where $\delta$ and $A$ are functions of $(t,r)$.
 Next, we assume that the wave maps are corotational, that is
\begin{equation}
F= F(t,r),\quad \Theta=\theta, \quad \Phi=\phi.
 \end{equation}
Eq.(3) reduces then to the single semilinear wave equation
\begin{equation}
\square_g F - \frac{\sin(2 F)}{r^2} = 0,
\end{equation}
where
\begin{equation}
\square_g = -e^{\delta} \partial_t(e^{\delta} A^{-1} \partial_t)
+\frac{e^{\delta}}{r^2} \partial_r(r^2 e^{-\delta} A\:
\partial_r),
\end{equation}
and the Einstein equations are
\begin{eqnarray}
{\partial_t A} &=& -2 \alpha\: r A (\partial_t F) (\partial_r F),
\\
{\partial_r \delta} &=& -\alpha\: r \left((\partial_r F)^2 +
A^{-2} e^{2\delta} (\partial_t F)^2 \right),
\\
\partial_r A &=& \frac{1-A}{r} - \alpha\: r \left( A (\partial_r F)^2 + A^{-1}
e^{2\delta} (\partial_t F)^2 + 2 \:\frac{\sin^2{\!F}}{r^2}\right).
\end{eqnarray}
These equations are invariant under dilations $(t,r) \rightarrow
(\lambda t, \lambda r)$ so it is natural to look for continuously
self-similar (CSS) solutions, that is solutions which are left
invariant by the action of the homothetic Killing vector $K=
t\partial_t+r\partial_r$. To study such solutions it is convenient
to use   similarity variables $\rho=r/(-t)$ and $\tau=-\ln(-t)$.
Then $K=-\partial_{\tau}$, so CSS solutions do not depend on
$\tau$. Assuming this and using an auxiliary function
$z=e^{\delta} \rho/A$, we reduce Eqs.(8-12)  to the system of
ordinary differential equations (where prime is $d/d\rho$)
\begin{equation}
F''+ \frac{2}{\rho} F' - \alpha(1+z^2)\rho {F'}^3  - \frac{\sin(2
F)}{A \rho^2(1-z^2)} = 0,
\end{equation}
\begin{eqnarray}
A'&=& - 2 \alpha \rho A {F'}^2,\\
\rho z' & =& z (1+\alpha (1-z^2) \rho^2 {F'}^2),\\
\rho A' &=& 1-A -\alpha \left(\rho^2 A (1+z^2) {F'}^2 + 2
\sin^2{F}\right).
\end{eqnarray}
The combination of (14) and (16)  yields the constraint
\begin{equation}
1-A  - 2 \alpha \sin^2{F} +\alpha A \rho^2 {F'}^2 (1-z^2) =0.
\end{equation}
In what follows, we will be solving Eqs.(13-15). The constraint
(17), which is the first integral of these equations, could be
used to eliminate $A$ from (13) but we find more convenient not to
do so. The system (13-15) has a fixed singularity at the center
$\rho=0$ and moving singularities at points where $z(\rho)=\pm 1$
and/or $A(\rho)=0$. In terms of the similarity coordinate $\rho$,
the metric (6) takes the  form
\begin{equation}
  ds^2 = A^{-1} (1-z^{-2})\:\rho^2  dt^2 + 2 A^{-1} t \rho\: dt d\rho
  + A^{-1} t^2 d\rho^2 + t^2 \rho^2 (d\theta^2+\sin^2{\!\theta}\: d\phi^2),
\end{equation}
hence the hypersurfaces $z=\pm 1$ are null (provided that $A>0$).
After~\cite{carsten} we will refer to these characteristic
hypersurfaces as to the self-similarity horizons (SSH). The first
$\rho_0$ where $z(\rho_0)=1$ is the locus of the past light cone
of the singularity at the origin $(t=0,r=0)$ and we will call it
the past SSH. We use the remaining coordinate freedom to locate it
at $\rho_0=1$, that is  $z(1)=1$. In the next section we will
analyze the system (13-15) below the past SSH, that is for $0 \leq
\rho \leq 1$.
\section{From the center to the past self-similarity horizon}

Expanding about $\rho=1$, we get a one-parameter family of
solutions of Eqs.(13-15)
\begin{equation}
F(\rho) \sim \frac{\pi}{2} + b (\rho-1), \quad z(\rho) \sim
1+(\rho-1), \quad A(\rho) \sim 1-2 \alpha - 2 \alpha (1-2 \alpha)
b^2 (\rho-1).
\end{equation}
 We shoot these initial data towards $\rho=0$ and adjust
the parameter $b$ so that the solution satisfies the regularity
condition at the center,
\begin{equation}
F(0)=0 , \quad z(0)=0, \quad A(0)=1.
\end{equation}
We find numerically that for each $\alpha<1/2$, there is an
infinite sequence $\{b_n\}$, $n=0,1,\dots$, such that the
corresponding solutions $(F_n,A_n,z_n)$ satisfy the boundary
conditions (19) and (20).
 The index $n$ denotes the number of
solutions of the equation $F_n(\rho)=\pi/2$ on the interval $0<
\rho <1$. In the limit $\alpha\rightarrow 0$, the $n$th solution
tends uniformly to the corresponding $\alpha=0$ solution. In the
limit $\alpha \rightarrow 1/2$, the parameters $b_n(\alpha)$
diverge to infinity indicating that regular solutions disappear
for $\alpha \geq 1/2$. Fig.~1 shows how the shooting parameters
$b_n$ depend on $\alpha$ for $n=0$ and $n=1$. In Fig.~2 we plot
the $n=0$ and $n=1$ solutions for $\alpha=0.1$.

To summarize, we claim that for each $\alpha<1/2$ there exists a
countable family of CSS solutions which are analytic at the center
and at the past SSH. The structure of these solutions is basically
the same as for $\alpha=0$. Qualitatively new things happen beyond
the past SSH and now we turn our attention to this much more
interesting region.
\section{Beyond the past self-similarity horizon}
For $\alpha=0$ it was easy to show that all solutions starting at
$\rho=1$ with initial values (19) remain regular for all $\rho>1$.
As we shall see below this is still true for  $\alpha>0$ provided
that $b$ is sufficiently small (note, however, that the metric
function $A(\rho)$ is monotone decreasing, so $A(\infty)<1$, which
means that the geometry is not asymptotically flat but conical).
However, if $b$ is large, the function
 $z(\rho)$ is not  monotone increasing for all $\rho>1$ and a "sonic point"
 develops for some
finite $\rho_B>1$, that is $\lim_{\rho\rightarrow \rho_B}
z(\rho)=1$.  We stress that this "sonic point" is not a SSH but an
apparent horizon  because $\lim_{\rho\rightarrow \rho_B}
A(\rho)=0$ as follows easily from (14) and (15). These two kinds
of behaviour are illustrated in Fig.~2.

Note that $\rho=\infty$ corresponds to the hypersurface
$(t=0,r>0)$ so in order to analyze the global behaviour of
solutions (for $t>0$) we need to go "beyond $\rho=\infty$". To
this end we define a new coordinate $x$ by
\begin{equation}
\frac{d}{dx}=\rho z \:\frac{d}{d \rho}, \qquad x(\rho=1)=0.
\end{equation}
We also define an auxiliary function $w(x)=1/z(\rho)$. In these
new variables, the past SSH where $w=1$ is at $x=0$, while the
future SSH (which nota bene is a Cauchy horizon) is located at
some $x_A>0$ where $w(x_A)=-1$.

In terms of $x$ and $w$, Eqs.(13-15) become autonomous (where now
prime is $d/dx$)
\begin{eqnarray}
w' &=& -1 +\alpha (1-w^2) {F'}^2,\\
A'&=& -2 \alpha A w {F'}^2, \\
(A F')'& =& \frac{\sin(2 F)}{w^2-1}.
\end{eqnarray}
The constraint (17) becomes
\begin{equation}
1-A  - 2 \alpha \sin^2{F} +\alpha A {F'}^2 (w^2-1) =0.
\end{equation}
 From (19) the initial conditions at $x=0$ are
\begin{equation}
F(x) \sim \frac{\pi}{2} + b x,\quad w(x) \sim 1 -x, \quad A(x)
\sim 1-2\alpha - 2 \alpha (1-2 \alpha) b^2 x.
\end{equation}
We already know from the previous section that for each
$\alpha<1/2$ there is an infinite sequence $\{b_n\}$ determining
solutions which are regular for all $x\leq 0$ (note that $\rho=0$
corresponds to $x=-\infty$). Although these solutions are
completely fixed at the past SSH by the requirement of analyticity
at the center, in order to understand their behaviour for $x>0$ it
is helpful to drop this requirement temporarily and consider
solutions that start from generic initial conditions (26). Of
course, such solutions in general are not analytic at the center.

 We first shall show that if $b$ is sufficiently small, then the
 solution
remains regular up to the future SSH. To see this, let
$f=(F-\pi/2)/b$. Then, in the limit $b \rightarrow 0$, Eqs.(22-24)
reduce to
\begin{eqnarray}
w' &=& -1 +\alpha (1-w^2) b^2 {f'}^2 \rightarrow -1,\\
A'&=& -2 \alpha A w\: b^2 {f'}^2 \rightarrow 0, \\
(A f')'& =& -\frac{\sin(2 b f)}{b (w^2-1)} \rightarrow -\frac{2
f}{w^2-1},
\end{eqnarray}
with the initial conditions
\begin{equation}
f(0)=0, \quad f'(0)=1,\quad w(0)= 1, \quad A(0)=1-2\alpha.
\end{equation}
The limiting  equations (27) and (28) are solved by $w=1-x$ and
$A=1-2\alpha$. Substituting these solutions into (29) we get the
equation
\begin{equation}
(1-2 \alpha) f''+ \frac{2 f}{x (x-2)} =0,
\end{equation}
whose solution is given by the hypergeometric function
 $f(x)=x\: {}_2\! F_1[(1-i\sqrt{7})/2,(1+i\sqrt{7})/2,2,x/2]$.
 Since solutions of
Eqs.(22-24) are continuous  in $b$ and $x$, by uniform continuity
 on compact
intervals, the solutions with sufficiently small $b$ will tend to
$w=-1$.

 Asymptotic analysis at the future SSH, $x_A$, yields the following
  leading order behaviour
 (where $y=x_A-x$)
\begin{equation}
w \sim -1+y, \quad A \sim A_0 - 2\alpha A_0 c^2 y \ln^2(y), \quad
F \sim F_0 - c y \ln(y),
\end{equation}
where $A_0=1-2\alpha \sin^2{F_0}$, $c=-\sin(2 F_0)/2 A_0$, and
$F_0$ is a free parameter. Thus, a solution that is analytic at
the past SSH, generically will be only $C^0$ at the future
SSH\footnote{If $F_0$ is an integer multiple of $\pi/2$, then the
singular log terms in (32) are absent and the solution is smooth
at the future SSH. One can show that this happens for discrete
values of $b$, which means that there exist solutions which are
analytic \emph{both} at the past and the future SSH. Of course,
for generic $\alpha$ these solutions are singular at the center.}.
Nevertheless, we checked that with the asymptotic behaviour (32)
all curvature invariants remain finite as $x \rightarrow x_A$. We
interpret this somewhat surprising fact as an indication that the
singularity at the origin is naked\footnote{A similar behaviour
for CSS solutions of the Einstein-axion-dilaton equations  was
observed by Eardley, Hirschmann, and Horne~\cite{ehh}.}.

Next, we shall show that solutions with large $b$ develop a
"sonic point" for some finite $x_B>0$, that is,
$\lim_{x\rightarrow x_B} w(x)=1$ and $\lim_{ x\rightarrow x_B}
A(x)=0$. This time, we define the variables
\begin{equation}
\xi=b^2 x, \quad h(\xi)=b^2 (1-w(x)), \quad s(\xi)=
 b(F(x)-\frac{\pi}{2}).
\end{equation}
Then, in the limit $b\rightarrow \infty$, Eqs.(22-24) reduce to
(where now prime is $d/d\xi$)
\begin{eqnarray}
h' &=& 1 -\alpha h \left(2-\frac{h}{b^2}\right) {s'}^2 \rightarrow
1-2\alpha h {s'}^2,\\
A'&=& -2 \alpha A \left(1-\frac{h}{b^2}\right) {s'}^2  \rightarrow
-2\alpha A {s'}^2,
\\
(A s')'& =& \frac{\sin(2 s/b)}{b h (2-h/b^2)} \rightarrow 0,
\end{eqnarray}
with the initial conditions
\begin{equation}
h(0)=0, \quad A(0)=1-2\alpha, \quad s(0)=0, \quad s'(0)=1.
\end{equation}
It follows from (36) and (37) that, in the limit $b \rightarrow
\infty$, $A s'=1-2 \alpha$. Plugging this into (34) and (35), and
using (37), we get the limiting solution
\begin{equation}
A(\xi)=(1-2 \alpha) \sqrt{1-4 \alpha \xi} \quad \mbox{and} \quad
h(\xi)= \frac{1}{2\alpha} \sqrt{1-4 \alpha \xi}\: (1-\sqrt{1-4
\alpha \xi}\:).
\end{equation}
This solution develops a "sonic point" at $\xi=1/4\alpha$. Again,
by uniform continuity on compact intervals, we conclude
 that for solutions
of Eqs.(22-24) with large $b$ (and nonzero $\alpha$), the function
 $w(x)$ attains a
minimum and then tends to $1$ at some $x \rightarrow x_B$, while
the function $A(x)$ drops to zero at $x_B$.
 The
leading order behaviour at the apparent horizon at $x_B$ is
\begin{equation}
w \sim 1 - a \sqrt{x_B-x}, \quad A \sim d \sqrt{x_B-x},\quad F
\sim F_B + \frac{1}{\sqrt{\alpha}} \sqrt{x_B-x},
\end{equation}
where the positive parameters $F_B,a,d$ are constrained by the
relationship $1- 2 \alpha \sin^2{F_B} - a d/2=0$. Substituting
this asymptotic behaviour into the metric (18), one can readily
verify that the hypersurface $x=x_B$ is spacelike.

Thus, solutions which start from the past SSH with initial
conditions (26), tend in finite "time" to $w=-1$ if $b$ is small,
or to $w=+1$ if $b$ is large. In what follows, we will refer to
these two behaviours as to type A and  type B solutions,
respectively. The solutions of type A obviously form an open set
and we believe (but have no proof yet) that the same is true for
the  solutions of type B. If so, there must exist solutions which
are not of type A or B, call them type C solutions. By definition,
type C solutions remain in the strip $w\in(-1,1)$ which implies
that they exist for all $x$ (because one can easily show that
solutions can go singular only if $w \rightarrow\pm 1$).

Our numerical analysis of the structure of types A, B, and C can
be summarized as follows\footnote{To simplify the exposition, we
are cheating at this point. We have evidence that for large values
of $\alpha$ ($\sim 0.42$) the transition between types A and B
occurs not at a single point but in a narrow interval
$(b_{min}^*,b_{max}^*)$. We suspect that this interval, which can
be thought of as the boundary between the basins of attractions of
type A and B behaviours, has a fractal structure. This intriguing
fact, suggesting perhaps a chaotic behaviour, is under
investigation and will be described  elsewhere.}. For a given
$\alpha$, there exists a critical value $b^*(\alpha)$ such that
solutions with $b<b^*$ are of type A, solutions with $b>b^*$ are
of type B, and the solution with $b=b^*$ is of type C. Thus, we
have a bistable behaviour with two generic final states A and B,
and the separatrix C.  Although the precisely critical initial
condition $b=b^*$  cannot be prepared numerically, in order to
figure out a behaviour of the type C solution it is sufficient
(due to continuous dependence  of solutions on initial conditions)
to determine the flow of nearly critical initial conditions
$b=b^*\pm \epsilon$. As is shown in Fig.~3, before the marginally
critical
 type A and B solutions tend to  $w=\pm 1$, they exhibit a long transient
 almost periodic
behaviour. This leads us to conjecture that the type C solution is
asymptotically periodic. The periodic intermediate attractor is
clearly seen on the $(w,A)$ plane.

 The complete discussion of periodic
solutions  would take us too far afield, so we postpone it to the
subsequent paper. Here, let us only point out that the behaviour
of CSS solutions outside the past SSH resembles some aspects of
the DSS type II critical behaviour at the threshold for black hole
formation. Actually, it can be interpreted as a poor ODE version
of this phenomenon. The type A and B solutions are the analogues
of dispersive and black-hole solutions, respectively. The type C
solution is the analogue of the critical evolution which tends
asymptotically to the critical periodic solution (which is the
analogue of the choptuon). We even have the analogue of the
black-hole-mass scaling. Namely, let $\lambda$ be the unstable
eigenvalue (that is $Re(\lambda)>0$) around the periodic solution.
Then the distance of a nearly critical flow (corresponding to
$b=b^*+\epsilon$) from the separatrix C is proportional to
$\epsilon \exp(Re(\lambda) x)$, which means that the length of the
transient periodic phase  is proportional to
$-\frac{1}{Re(\lambda)} \ln|\epsilon|$. The locus of the future
SSH  $x_A$ (for the marginally critical type A solutions) and the
locus of an apparent horizon $x_B$ (for the  marginally critical
type B solutions) scale in the same manner (see Fig.~4). This
leading order scaling law is decorated in the next order by
periodic wiggles with period $Re(\lambda) T$. The origin of these
wiggles is basically the same as in the case of  black-hole mass
scaling in the type II DSS critical collapse~\cite{carsten}. In
passing, we remark that
 the periodic solutions are
nonperturbative in $\alpha$ -- in the limit $\alpha \rightarrow
0$, the period and the amplitude of oscillations go to zero.

Now, we return to the countable family of CSS solutions
constructed in section~3 and finally determine their extension
beyond the past SSH. As follows from the above discussion, for a
given $\alpha$, the behaviour of the $n$th solution outside the
past SSH depends on whether $b_n$ is smaller or larger than $b^*$:
the solutions with $b_n<b^*$ are of type A (and therefore regular
in our terminology), while the solutions with $b_n>b^*$ are of
type B. The function $b^*(\alpha)$ is monotone decreasing and
$b^*(\alpha) \rightarrow \infty$ as $\alpha \rightarrow 0$, while
the functions $b_n(\alpha)$ are monotone increasing, hence we
obtain an infinite sequence $\{\alpha_n\}$ defined by the
intersections $b^*(\alpha_n)=b_n(\alpha_n)$ (see Fig.~1). Since
$b_{n+1}(\alpha)<b_n(\alpha)$ and $b_n(\alpha)\rightarrow \infty$
as $\alpha\rightarrow 1/2$, the sequence $\{\alpha_n\}$ is
increasing and bounded above by $1/2$.
 Therefore, we conclude that the $n$th CSS solution from section 3
 is regular iff $\alpha<\alpha_n$. The first two values of
 $\alpha_n$ are $\alpha_0 \approx 0.0688$ and $\alpha_1 \approx
 0.1588$.
\section{Final remarks}
The role of the CSS solutions considered in this paper in the
dynamical evolution depends crucially on their stability with
respect to small perturbations. Although we have not performed the
stability analysis, it seems plausible that  for sufficiently
small $\alpha$ the stability properties are the same as for
$\alpha=0$, that is the $n$th solution has exactly $n$ unstable
modes. Speculations that follow are based on this assumption. We
expect that for $\alpha<\alpha_0$ the situation is qualitatively
the same as in the flat spacetime, that is the $n=0$ solution
determines the universal asymptotics of a naked singularity
formation and the $n=1$ solution sits at the threshold of this
process. For $\alpha>\alpha_0$ the singularity of the $n=0$
solution is covered by a horizon so generic naked singularities
presumably disappear -- nota bene this could be interpreted as a
gravitational desingularization phenomenon. In the range
$\alpha\in (\alpha_0,\alpha_1)$ we expect the  type II threshold
behaviour with the $n=1$ CSS critical solution. What happens for
$\alpha>\alpha_1 \approx 0.1518$? Since the $n=1$ solution is not
regular, it cannot be a bona fide critical solution (it can still
appear as a local intermediate attractor, though). HLPTA showed
in~\cite{husa1} that for large values of $\alpha$ the critical
behaviour is DSS and the echoing period $\Delta$ increases rapidly
as $\alpha$ tends to $0.18$ from above (see Fig.~4
in~\cite{husa1}). Moreover, they noticed small deviations from
exact DSS at this smallest $\alpha$ value for which they observed
DSS critical collapse. These facts led them to conjecture that a
critical solution ceases to be DSS for still smaller values of
$\alpha$. This conjecture fits nicely with our conjecture that a
critical solution ceases to be CSS for $\alpha>\alpha_1$. Taken
together, these conjectures suggest that the transition from CSS
to DSS critical behaviour is not sharp but rather has a crossover
character.  HLPTA are currently running the computer simulations
of the critical behaviour for the coupling constants
$\alpha<0.18$, so the above speculations should be verified (or
falsified) soon. Hopefully, the numerical phenomenology will also
help understand a yet unknown mechanism of the CSS/DSS changeover.
\section*{Acknowledgments}
 The research of PB was supported in part
by the KBN grant 2 P03B 010 16. PB thanks Peter Aichelburg and his
coworkers for discussions and sharing  their numerical results
with him.

 \newpage
 \begin{figure}
\centering
\includegraphics[width=\textwidth]{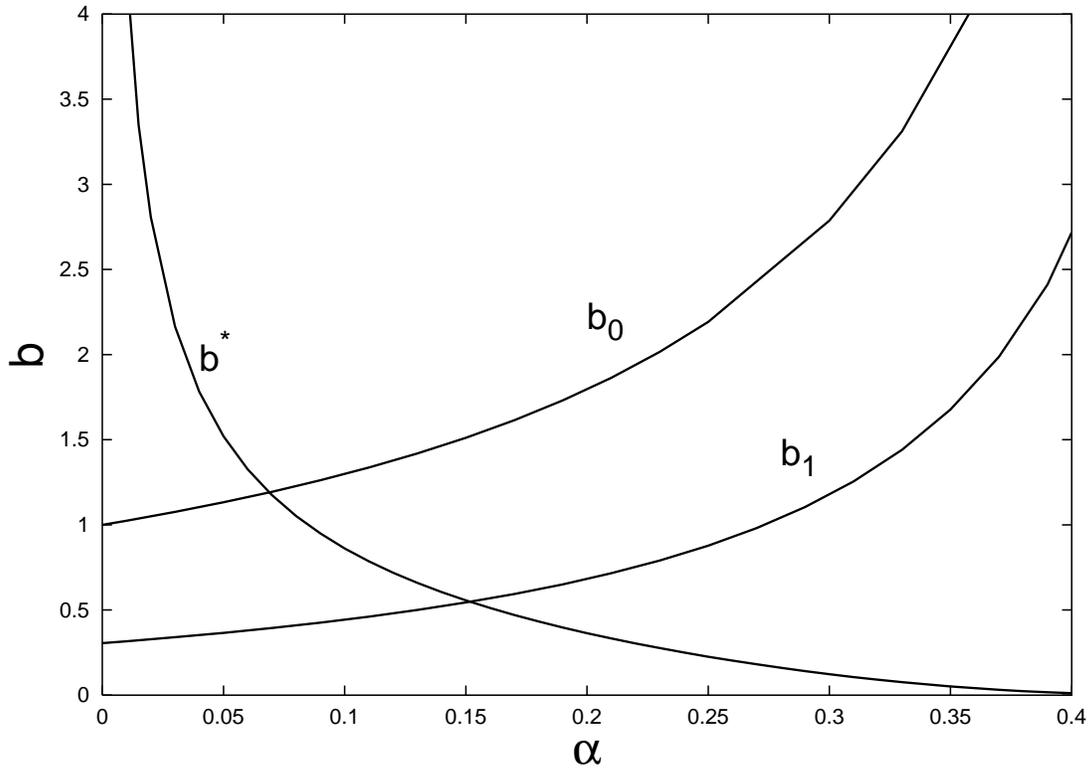}
\caption{The shooting parameters $b_n(\alpha)$ for $n=0$ and
$n=1$. The critical curve $b^*(\alpha)$ separates solutions that
have an apparent horizon beyond the past SSH from those that do
not. The intersections of the curve $b^*(\alpha)$ with the curves
$b_0(\alpha)$  and $b_1(\alpha)$ determine the critical values of
the coupling constant $\alpha_0 \approx 0.0688$ and  $\alpha_1
\approx 0.1518$.}
\end{figure}
\begin{figure}
 \centering
\includegraphics[width=0.7\textwidth]{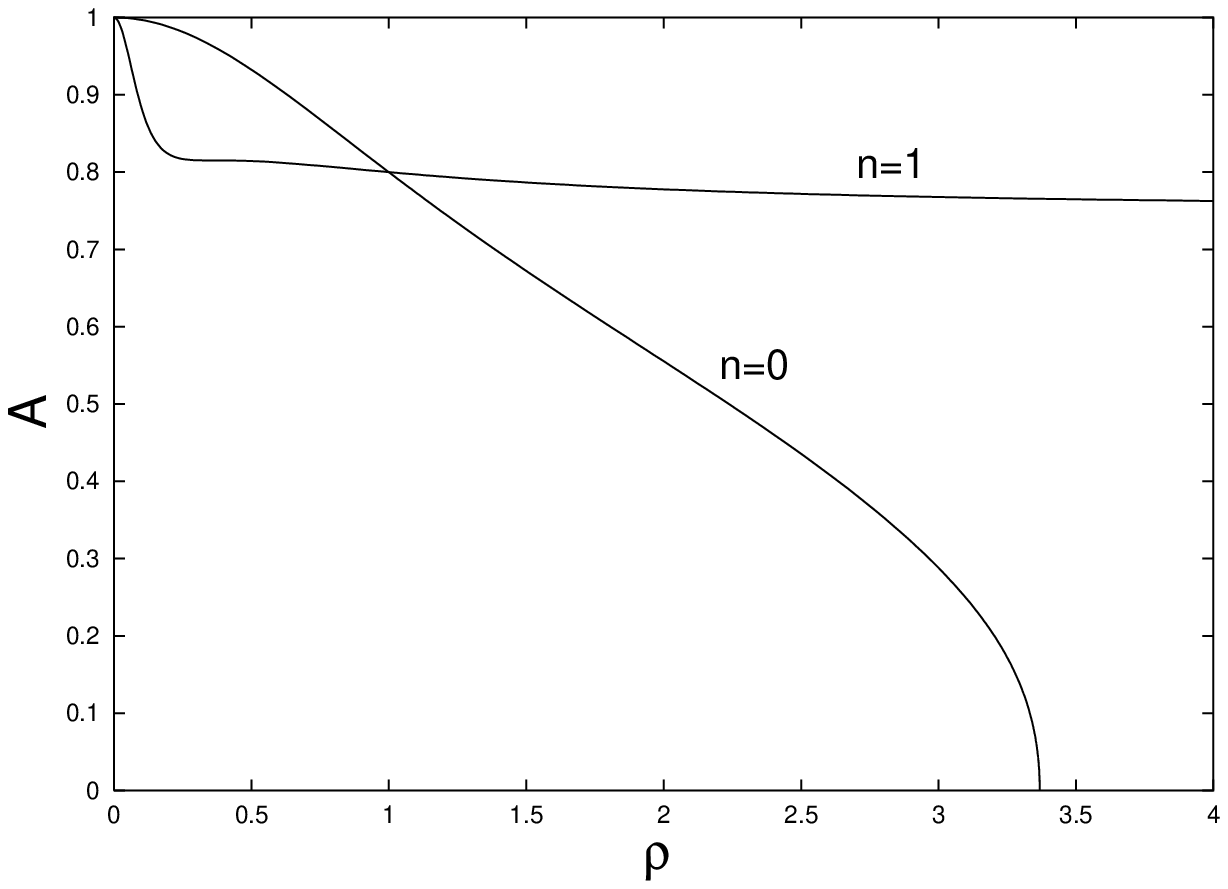}\\
\vspace{0.1in}%
\centering
\includegraphics[width=0.7\textwidth]{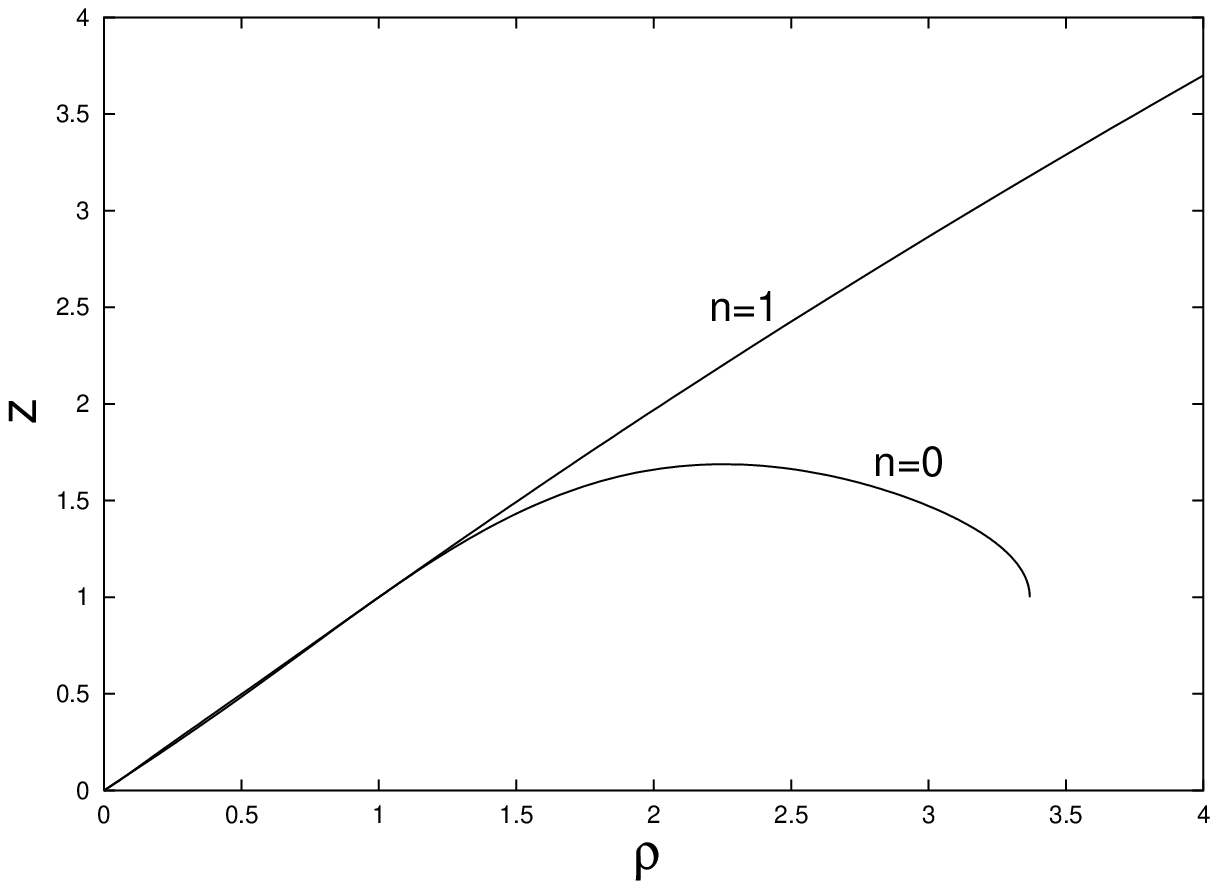}\\
\vspace{0.1in}%
\centering
\includegraphics[width=0.7 \textwidth]{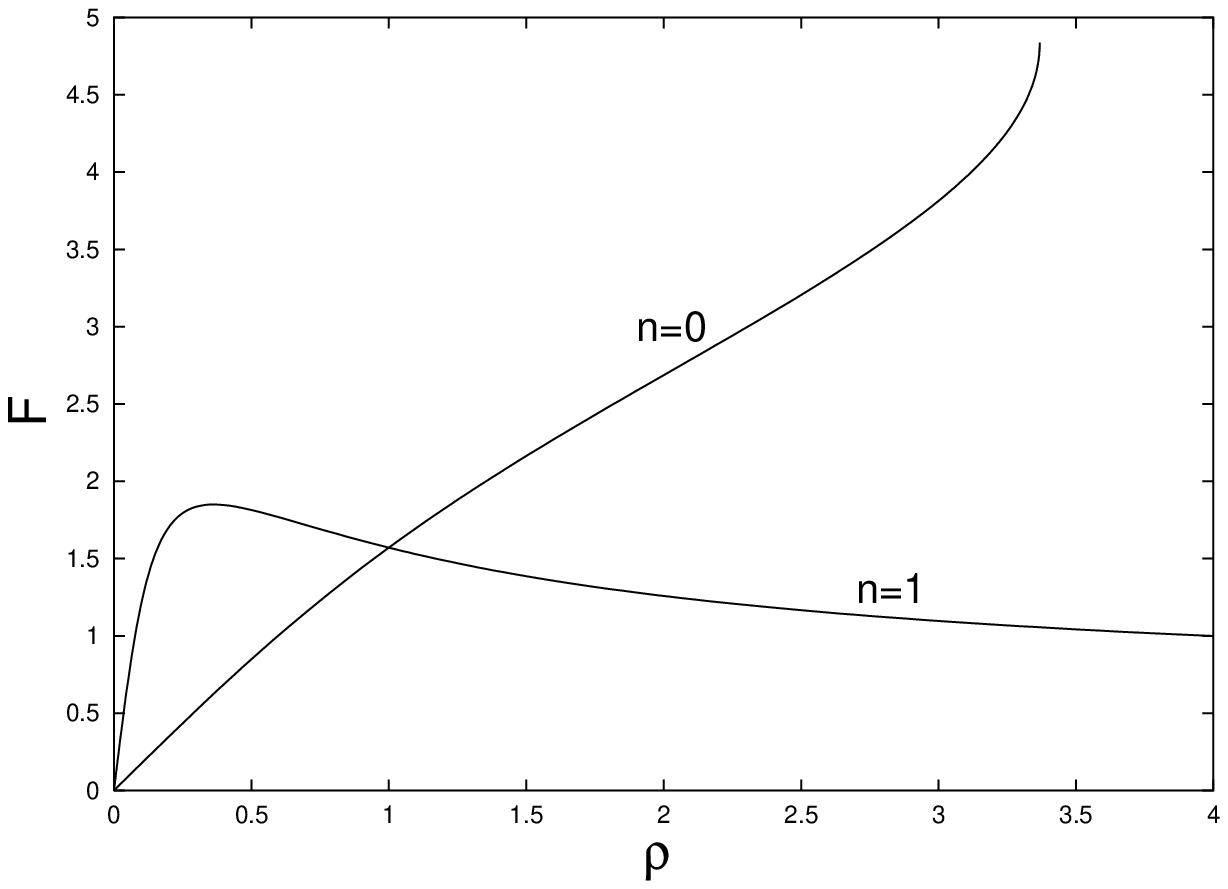}%
 \caption{The $n=0$ and $n=1$ CSS solutions for $\alpha=0.1$. Since
 $\alpha_0<0.1<\alpha_1$, the $n=1$ solution exists for all $\rho>1$, whereas
 the $n=0$ solution develops an apparent horizon at $\rho_B \approx 3.3683$.}
\end{figure}
\begin{figure}
 \centering
\includegraphics[width=0.7\textwidth]{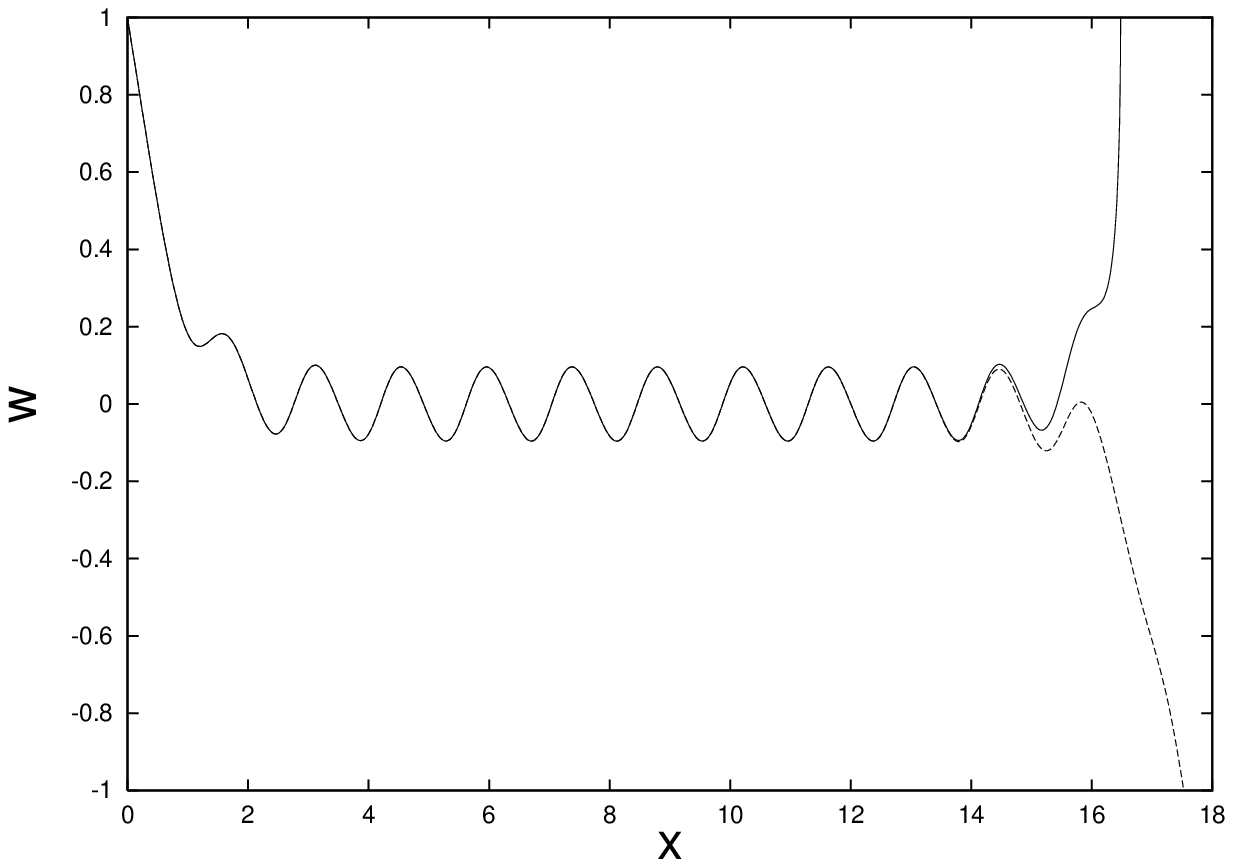}\\
\vspace{0.1in}%
\centering
\includegraphics[width=0.7\textwidth]{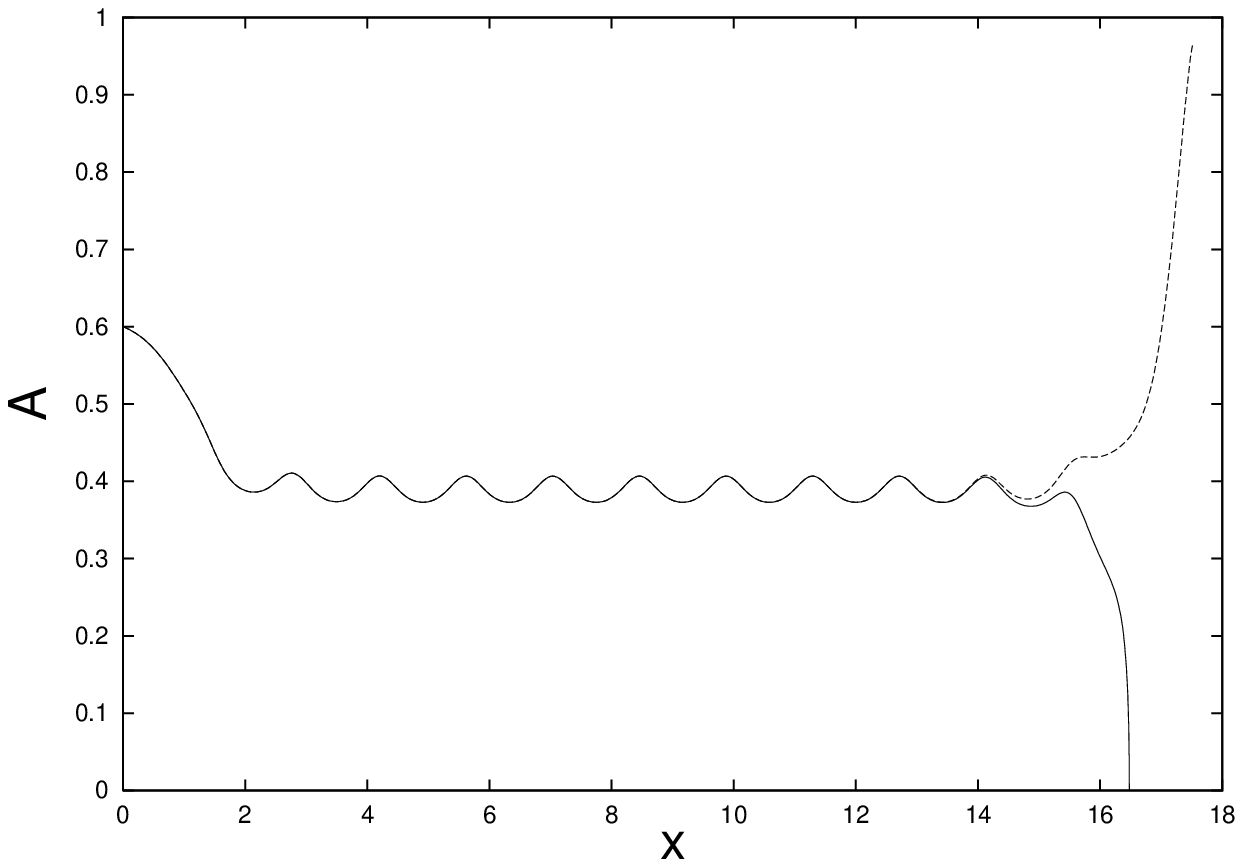}\\
\vspace{0.1in}%
\centering
\includegraphics[width=0.7 \textwidth]{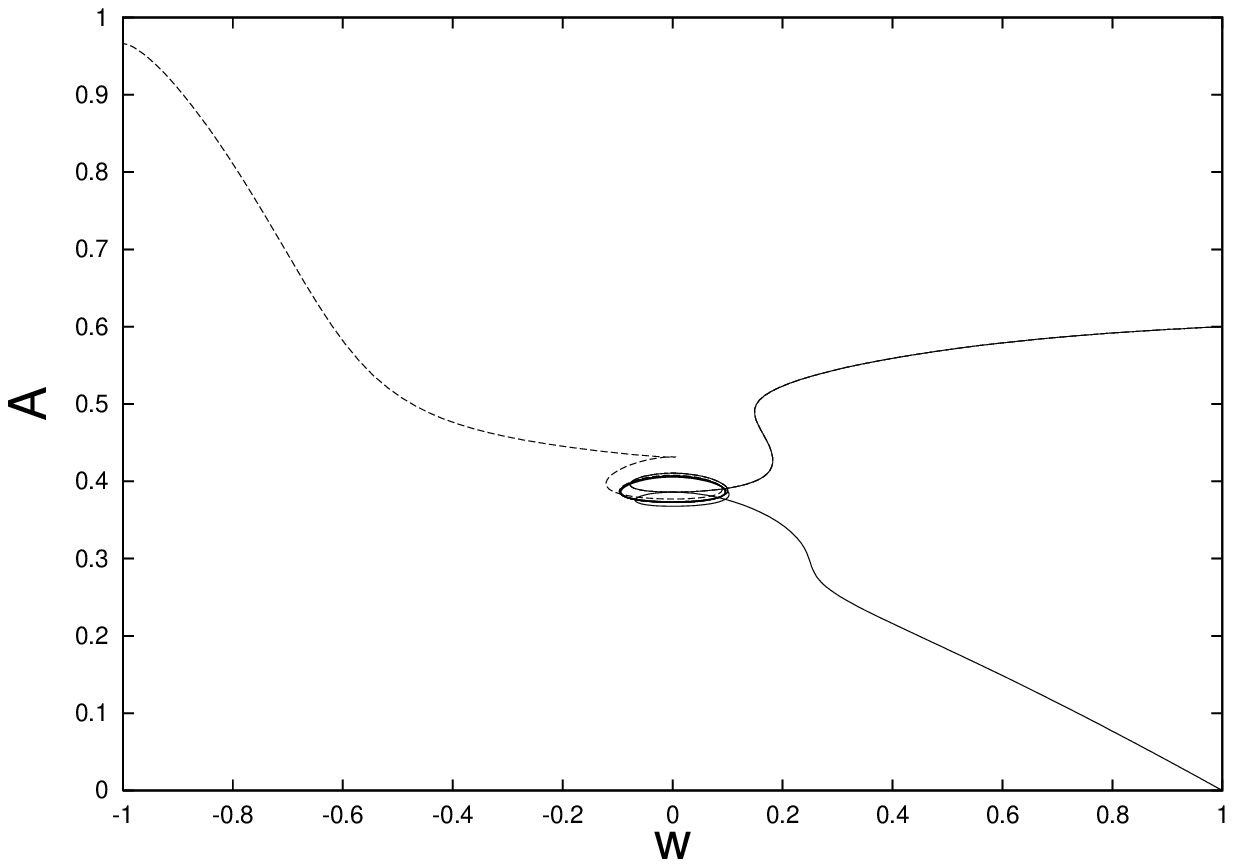}%
 \caption{The metric functions $w$ and $A$ for two
 nearly critical solutions for $\alpha=0.2$:
  type A solution with $b=0.36425022396604$ (dashed line) and type B
  solution
  with  $b=0.36425022396605$ (solid line). For intermediate values of $x$
  the solutions are almost periodic with the period $T\approx 1.418$. The
   intermediate attractor is seen on the $(w,A)$ plane as an unstable limit
    cycle.}
\end{figure}
 \begin{figure}
\centering
\includegraphics[width=\textwidth]{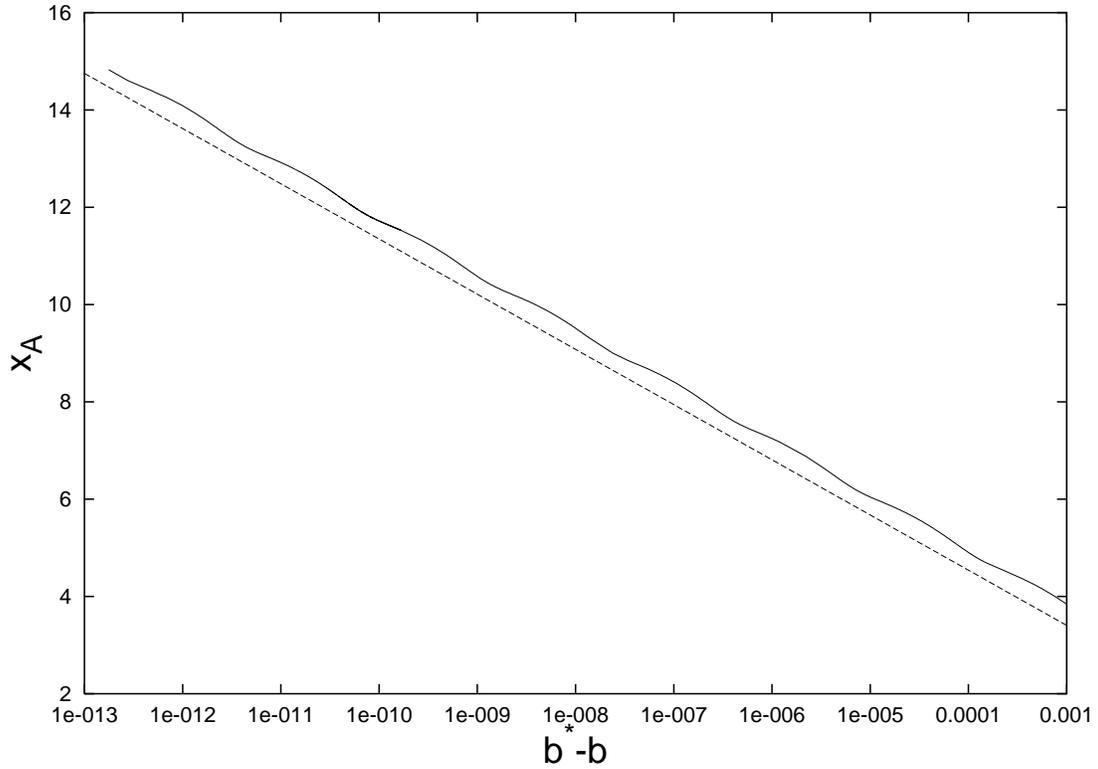}
\caption{The locus of the future SSH, $x_A$, against the
logarithmic distance from the critical value, $\ln(b^*-b)$, for
$\alpha=0.2$ is plotted as the solid line. The dashed line
(shifted down for better comparison) shows the least-square fit of
the leading order behaviour $x_A = -\frac{1}{Re(\lambda)}
\ln(b^*-b)$ with $Re(\lambda) \approx 2.029$. The wiggles
superimposed on this straight line are the imprint of periodicity
of the intermediate attractor. The numerically computed period of
the wiggles is approximately equal to $2.887$ which agrees with
the theoretically predicted value $Re(\lambda)\: T$.}.
\end{figure}
\end{document}